\documentstyle[aps,multicol]{revtex}
\draft

\begin{document}
\newcommand{\smallfrac}[2]{\mbox{$\frac{#1}{#2}$}}

\title{Feedback-control of quantum systems using continuous
  state-estimation} 
\author{A. C. Doherty and K. Jacobs}
\address{Department of Physics, University of Auckland, Private Bag 
92019, Auckland, New Zealand.}
\maketitle

\begin{abstract}
We present a formulation of feedback in quantum systems in which the best
estimates of the dynamical variables are obtained continuously from the
measurement record, and fed back to control the system. We apply this
method to the problem of cooling and confining a single quantum degree of
freedom, and compare it to current schemes in which the measurement
signal is fed back directly in the manner usually considered in existing
treatments of quantum feedback. Direct feedback may be combined with 
feedback by estimation, and the resulting combination, performed on a 
linear system, is closely analogous to classical LQG control theory with
residual feedback.

\end{abstract}

\pacs{42.50.Lc,03.65.Bz,42.50.Ct}
\begin{multicols}{2}

\section{Introduction}
The continuous measurement of quantum systems has been a topic of
considerable activity in recent 
years~\cite{cm0,c0b,cm1,cm2,cm3,cm4,cm6,cm7,WMhom,GWDR}, 
and is particularly relevant at this time because experimental 
technology is now at the point where {\em individual} quantum systems 
can be monitored continuously~\cite{Mabuchi}. With these developments 
it should be possible in the near future to control quantum systems 
in real time by using the results of the measurement in the process of 
continuous feedback.

A theory describing the dynamics that results from feeding back the 
measurement signal (usually a photocurrent) at each instant to control
the  Hamiltonian of a quantum system has been developed by Wiseman and 
Milburn~\cite{qfb}. They have shown how to derive the resulting Stochastic
Master Equation (SME) for the conditioned evolution, and the corresponding
unconditioned master equation, both of which are Markovian in the limit of
instantaneous feedback. This kind of feedback has already been used to
reduce laser noise below the shot-noise level~\cite{fbQND}.

However, there are many ways in which the measurement signal may be fed
back to affect the system. In general, at a given time, any integral of
the measurement record up until that time may be used to alter the system
Hamiltonian and affect the dynamics. This leads, however, to a
non-Markovian master equation, and the resulting dynamics cannot
therefore be easily investigated, and, more importantly, understood.
Nevertheless, as we shall examine in this paper, certain integrals of the
measurement record provide specific information, such as the best
estimates of dynamical variables. These may be fed back to alter the
system evolution in a desired way, and while the unconditioned evolution
of the system is no longer Markovian, simple equations may be derived for 
the selective evolution of system variables, and 
correspondingly simple non-linear (but Markovian) SME's describe the 
evolution of the system in the limit of
instantaneous feedback. This approach to quantum feedback has close
analogies to that used in classical control theory, in particular that
control is broken down into a state-estimation step and a feedback step.
Because of this, classical results regarding the design of feedback loops
can be applied, opening up new possibilities for controlling quantum
systems.

In this paper we consider a single quantum degree of freedom, which 
could be, for example, a trapped atom~\cite{Doh1}, ion~\cite{ion} or a
moving mirror forming one end of an optical cavity~\cite{BJK}, subjected
to continuous position observation. Naturally the continuous observation
of position actually corresponds to a continuous {\em joint measurement}
of both position and momentum, because momentum information is implicit in
the observed change in the position over time. We show how the best
estimate of both the position and momentum at each point in time may be
obtained from an integral of the measurement signal when the initial
state of the system is Gaussian.

We examine the dynamics which results from using the best estimates of 
the system variables in a feedback loop, and in particular 
investigate cooling and confinement using this mechanism. We also apply
the Wiseman-Milburn  direct feedback theory to investigate the
implementation of cooling and confinement by feeding back the measurement
signal at each time, a technique which has been considered by
Dunningham {\em et al.}~\cite{Dun} and Mancini {\em et
al.}~\cite{mancini}, and contrast this with the method involving
estimation.

In the next section we review briefly how the SME describing a continuous 
measurement of position results from
real physical measurement schemes. In section~\ref{solM} we review the
solution  of this master equation, which may be obtained in a simple
manner for initially Gaussian states, and give the integrals that are
required to measure both position and momentum (ie. to obtain the best
estimates of position and momentum at each time). We then examine the
dynamical equations which result from using the best estimates for the
purposes of feedback, and present the classical control theory which may
applied to this quantum feedback process due to an equivalence with
classical estimation theory. In section~\ref{fb1} we consider the problem
of cooling and confinement using feedback. We apply both feedback by
estimation and direct feedback to this problem.
Section~\ref{conc} concludes.

\section{Continuous position measurements}\label{cms}
\subsection{Two physical position measurement systems}
A continuous measurement of the position of a macroscopic object may be 
obtained by observing continuously the phase of a light beam reflected
from it. If we allow the object in question to form one end-mirror in an 
optical cavity, then in the limit in which one of the cavity mirrors is
very lossy (the bad-cavity limit), the phase of the light output from the
cavity provides a continuous measurement of the position of the moving 
mirror, since the light spends little time in the cavity~\cite{MJW,JTWS}. 
This is a simple way to treat position measurement by light reflection, 
and what is more, the position of a single atom may also be monitored 
continuously in the same manner. To monitor the position of a single 
atom, the atom is allowed to interact off-resonantly with the optical 
cavity mode, and this interaction is such that the atom generates a phase 
shift of the output light in a manner similar to the moving end-mirror. 
We now examine briefly these two situations, and derive the SME 
describing the measured systems.

The Hamiltonian describing an optical cavity in which one of the mirrors
is free to move along the axis of the cavity is~\cite{BJK}
\begin{equation}
H = H_m - \hbar g_m a^\dagger a x + H_d
\end{equation}
where $a$ is the annihilation operator for the cavity mode, 
$g_m=\omega_0/L$ is the coupling constant describing the interaction
between the cavity mode and the moving mirror (in which $\omega_0$ is
the mode frequency, and $L$ is the length of the cavity), $H_m$ is the
Hamiltonian for the mechanical motion of the mirror, and $H_d$ describes
the coherent driving of the cavity mode. Note that we have moved into the
interaction picture with respect to the free Hamiltonian of the cavity
mode. In deriving this Hamiltonian it is assumed that the cavity mode
follows the motion of the mirror adiabatically, and in particular that
the change in the cavity length due to the motion of the mirror is small
compared to the cavity length itself~\cite{Pace}. That is, $|\langle
x\rangle| \ll L$.  One of the end-mirrors is chosen to be lossy so as to
provide output coupling, and the cavity is driven through this mirror.
The part of the Hamiltonian which describes coherent driving of the
cavity is given by
\begin{equation}
H_d = i\hbar E(a - a^\dagger),
\end{equation}
where $E$ is related to the laser power $P$ by  $E=\sqrt{\gamma
P/(\hbar\omega_0)}$, and $\gamma$ is the decay rate of the cavity due to
the output coupling mirror~\cite{JTWS}.

We see that the Hamiltonian describing the interaction between the
mirror and the cavity field is of the form $a^\dagger a x$. This is
exactly what we need in order to obtain a continuous position
measurement by monitoring the phase of the output light. This is because
$a^\dagger a$ is the generator of a phase shift for the light, and
therefor a Hamiltonian of this form produces a phase shift proportional
to the position of the mirror, which is exactly what is required. 

The Hamiltonian describing the off-resonant interaction between a
two-level atom, and an optical cavity in which it is trapped,
is~\cite{WandM}
\begin{equation}
H = H_a - \hbar \frac{g_0^2}{\Delta} a^\dagger a \cos^2(k_0 x) + H_d
\end{equation}
where  $k_0=\omega_0/c$ is the wavenumber of the cavity mode, $\Delta$ is 
the detuning between the cavity mode and the two level atom
($\Delta=\omega_a - \omega_0$, where $\omega_a$ is the frequency of the
atomic transition), $x$ is the  atomic position operator, $g_0$ is
the cavity-QED coupling constant giving the strength of the interaction
between the cavity mode and the atom and $H_a$ is the
Hamiltonian for the mechanical motion of the atom.  We will assume the
atom to be harmonically trapped, which might be achieved by using a
second light field~\cite{Mabuchi}, or by ion tapping~\cite{trapcav}. 

To obtain a continuous position measurement by monitoring the phase of 
the output light, we require the interaction of the atomic motion and the
cavity mode to be of the same form as that for the mirror. To achieve
this we need simply ensure that the atom is trapped in a region small
compared to the wavelength of the light, about a region halfway between a
node and an anti-node so that we may approximate $\cos^2(k_0 x) =
\cos^2(k_0 x_0 + k_0x') \approx 1/2 + k_0x'$. Renaming $x'$ as $x$ 
(merely a shift in which the resulting extra term in the Hamiltonian is
unimportant), we obtain the correct interaction Hamiltonian.

To realize a position measurement the phase quadrature of the output 
light must be monitored, and we choose homodyne detection since it 
provides the simplest treatment. Performing homodyne detection of the 
phase quadrature with a detector efficiency $\eta$, the SME describing 
the evolution of
the system conditioned on the continuous measurement record 
is~\cite{cm7,WMhom} (see also \cite{cm1,cm2,cm4}),
\begin{equation}
 d\rho_{\mbox{\scriptsize c}} = -\frac{i}{\hbar}
[H,\rho_{\mbox{\scriptsize c}}] dt + \gamma {\cal D}[a]
\rho_{\mbox{\scriptsize c}} dt +
\sqrt{\eta\gamma} {\cal H}[-ia] \rho_{\mbox{\scriptsize c}}  dW,
\end{equation}
where $\rho_{\mbox{\scriptsize c}}$ is the system density matrix 
conditioned on the measurement record, $dW$ is the Wiener increment,
satisfying the Ito calculus relation $(dW)^2=dt$, and the superoperators
${\cal D}$ and ${\cal H}$ are given by
\begin{eqnarray}
2{\cal D}[c] \rho_{\mbox{\scriptsize c}} & = & 2 c \rho_{\mbox{\scriptsize
c}} c^\dagger - c^\dagger c \rho_{\mbox{\scriptsize c}} -
\rho_{\mbox{\scriptsize c}} c^\dagger c , \\ {\cal H}[c]
\rho_{\mbox{\scriptsize c}} & = & c \rho_{\mbox{\scriptsize c}} +
\rho_{\mbox{\scriptsize c}} c^\dagger - \mbox{Tr}[c
\rho_{\mbox{\scriptsize c}} + \rho_{\mbox{\scriptsize c}} c^\dagger ]
\rho_{\mbox{\scriptsize c}} ,
\end{eqnarray}
for an arbitrary operator $c$. 

\subsection{Adiabatic elimination of cavity dynamics}
We are interested only in the dynamics of the atom (or 
equivalently the mirror), and we are also interested purely in the
bad-cavity limit (large $\gamma$) which corresponds to good position
measurement . In this limit, due directly to the
high cavity damping rate, the cavity mode is slaved to the atom
dynamics, and can therefore be adiabatically eliminated to obtain a
SME purely for the atom.  To do this we proceed by following essentially 
the treatment in reference~\cite{WMhom}. 

Noting first that in the absence of the 
interaction with the atom, the steady state of the cavity mode is the 
coherent state $|\alpha=-2E/\gamma\rangle$, we transform the system using 
\begin{equation}
\rho_{\mbox{\scriptsize c}}' = D(-\alpha) \rho_{\mbox{\scriptsize c}}
D^\dagger (-\alpha)
\end{equation}
where $D(\alpha)$ is the displacement operator, such that $D(\alpha) 
|0\rangle = |\alpha\rangle$~\cite{WandM}. In this `displacement
picture', the steady-state of the cavity is now close to the vacuum,
with the SME being
\begin{eqnarray}
  d\rho_{\mbox{\scriptsize c}}' & =&  -\frac{i}{\hbar}
[H_{\mbox{\scriptsize m}} - \hbar g (a^\dagger a + \alpha (a+a^\dagger) 
 + |\alpha|^2 ) x ,\rho_{\mbox{\scriptsize c}}'] dt \nonumber \\
 & & + \gamma {\cal D}[a] \rho_{\mbox{\scriptsize c}'} dt +
\sqrt{\eta\gamma} {\cal H}[-ia] \rho_{\mbox{\scriptsize c}'}  dW. 
\end{eqnarray} 

The regime required for adiabatic elimination is
\begin{equation}
  \left|\frac{\langle H_m\rangle}{\gamma}\right| \sim  
   \frac{g (|\alpha|^2 + 1) |\langle x\rangle |}{\gamma} 
  = \epsilon \ll 1 ,
\label{assump1}
\end{equation}
where $\epsilon$ will be our small parameter governing the approximation. 
Here $g=g_{\mbox{\scriptsize m}}$ for the case of the mirror, or 
$g=k_0 g_0^2/\Delta$ for the atom. Relation (\ref{assump1}) also
implies that $(g \alpha |\langle x\rangle |/\gamma) < \epsilon$, so
this quantity also serves as the small parameter. To proceed we assume
that the elements of the cavity mode density matrix in the number
basis, $\rho'^{nm}_{\mbox{\scriptsize c}}$, scale with the small
parameter $\epsilon$ as $\rho'^{nm}_{\mbox{\scriptsize c}} \propto  
\epsilon^{\left( n+m\right)}$, and we will show that this is 
consistent with the regime (\ref{assump1}). Under this assumption, the 
state of the cavity+(atom/mirror) may then be expanded up to second
order in 
$\epsilon$ as
\begin{eqnarray}
\rho_{\mbox{\scriptsize c}}' & = &
\;\;\; \rho^{\mbox{\scriptsize a}}_{\mbox{\scriptsize 00}}
|0\rangle\langle 0| + (\rho^{\mbox{\scriptsize a}}_{\mbox{\scriptsize
10}}  |1\rangle\langle 0| + \mbox{H.c.}) \nonumber \\
& & + \rho^{\mbox{\scriptsize
a}}_{\mbox{\scriptsize 11}}  |1\rangle\langle 1| +
(\rho^{\mbox{\scriptsize a}}_{\mbox{\scriptsize 20}}  |2\rangle\langle 0|
+ \mbox{H.c.}) + O(\epsilon^3),
\end{eqnarray}
so that
\begin{equation}
\rho_{\mbox{\scriptsize a}} = \mbox{Tr}_{\mbox{\scriptsize c}}
[\rho_{\mbox{\scriptsize c}}'] =
\rho^{\mbox{\scriptsize a}}_{\mbox{\scriptsize 00}}  +
\rho^{\mbox{\scriptsize a}}_{\mbox{\scriptsize 11}}  + O(\epsilon^3).
\end{equation}
where $\mbox{Tr}_{\mbox{\scriptsize c}}$ represents a trace over the 
cavity mode. 

The equations of motion for the density matrix elements 
$\rho^{\mbox{\scriptsize a}}_{\mbox{\scriptsize ij}}$ may then be 
obtained from the master equation for $\rho_{\mbox{\scriptsize c}}'$,
giving
\begin{eqnarray}
 d \rho^{\mbox{\scriptsize a}}_{\mbox{\scriptsize 00}} & = & 
{\cal L}_{\mbox{\scriptsize m}}^0 
\rho^{\mbox{\scriptsize a}}_{\mbox{\scriptsize 00}} \; dt  + ig\alpha ( x 
\rho^{\mbox{\scriptsize a}}_{\mbox{\scriptsize 10}} - 
\rho^{\mbox{\scriptsize a}\dagger}_{\mbox{\scriptsize 10}} x ) dt +
\gamma 
\rho^{\mbox{\scriptsize a}}_{\mbox{\scriptsize 11}} dt \nonumber \\
 & & -i \sqrt{\eta\gamma}
 (\rho^{\mbox{\scriptsize a}}_{\mbox{\scriptsize 10}} -
 \rho^{\mbox{\scriptsize a}\dagger}_{\mbox{\scriptsize 10}}
 -\mbox{Tr}[\rho^{\mbox{\scriptsize a}}_{\mbox{\scriptsize 10}} -
 \rho^{\mbox{\scriptsize a}\dagger}_{\mbox{\scriptsize 10}}] 
\rho^{\mbox{\scriptsize a}}_{\mbox{\scriptsize 00}}) dW \nonumber \\
 & & + O(\epsilon^3) , 
 \label{eq00} \\
 d \rho^{\mbox{\scriptsize a}}_{\mbox{\scriptsize 10}} & = & 
{\cal L}_{\mbox{\scriptsize m}}^0 
\rho^{\mbox{\scriptsize a}}_{\mbox{\scriptsize 10}} \; dt - 
\smallfrac{\gamma}{2} 
\rho^{\mbox{\scriptsize a}}_{\mbox{\scriptsize 10}} dt \nonumber \\
 & & + ig [ x (\alpha 
\rho^{\mbox{\scriptsize a}}_{\mbox{\scriptsize 00}} + 
  \rho^{\mbox{\scriptsize a}}_{\mbox{\scriptsize 10}} + 
\sqrt{2}\alpha \rho^{\mbox{\scriptsize a}}_{\mbox{\scriptsize 20}}) - 
\alpha \rho^{\mbox{\scriptsize a}}_{\mbox{\scriptsize 11}} x ] 
dt \nonumber \\
 & & -i \sqrt{\eta\gamma} (\sqrt{2} \rho^{\mbox{\scriptsize
 a}}_{\mbox{\scriptsize 20}} 
 - \rho^{\mbox{\scriptsize
 a}}_{\mbox{\scriptsize 11}}   - 
 \mbox{Tr}[ \rho^{\mbox{\scriptsize a}}_{\mbox{\scriptsize 10}}
  - \rho^{\mbox{\scriptsize a}\dagger}_{\mbox{\scriptsize 10}} ] 
\rho^{\mbox{\scriptsize a}}_{\mbox{\scriptsize 10}}) dW, \nonumber \\
 & & + O(\epsilon^3) \\
 d \rho^{\mbox{\scriptsize a}}_{\mbox{\scriptsize 11}} & = & 
{\cal L}_{\mbox{\scriptsize m}}^1 
\rho^{\mbox{\scriptsize a}}_{\mbox{\scriptsize 11}} \; dt  
+ ig\alpha ( x
\rho^{\mbox{\scriptsize a}\dagger}_{\mbox{\scriptsize 10}} - 
\rho^{\mbox{\scriptsize a}}_{\mbox{\scriptsize 10}} x ) dt - \gamma 
\rho^{\mbox{\scriptsize a}}_{\mbox{\scriptsize 11}} dt \nonumber \\
 & & + i \sqrt{\eta\gamma}
 \mbox{Tr}[\rho^{\mbox{\scriptsize a}}_{\mbox{\scriptsize 10}} -
 \rho^{\mbox{\scriptsize a}\dagger}_{\mbox{\scriptsize 10}}] 
 \rho^{\mbox{\scriptsize a}}_{\mbox{\scriptsize 11}} dW + O(\epsilon^3) , 
 \label{eq11} \\
 d \rho^{\mbox{\scriptsize a}}_{\mbox{\scriptsize 20}} & = & 
{\cal L}_{\mbox{\scriptsize m}}^0 
\rho^{\mbox{\scriptsize a}}_{\mbox{\scriptsize 20}} \; dt - 
\gamma \rho^{\mbox{\scriptsize a}}_{\mbox{\scriptsize 20}} 
 dt + ig x ( 2 \rho^{\mbox{\scriptsize a}}_{\mbox{\scriptsize 20}}
 + \sqrt{2}\alpha \rho^{\mbox{\scriptsize a}}_{\mbox{\scriptsize 10}}) dt 
 \nonumber \\
 & & + i\sqrt{\eta\gamma} \mbox{Tr}[ 
\rho^{\mbox{\scriptsize a}}_{\mbox{\scriptsize 10}} -
 \rho^{\mbox{\scriptsize a}\dagger}_{\mbox{\scriptsize 10}} ] 
 \rho^{\mbox{\scriptsize a}}_{\mbox{\scriptsize 20}} dW 
+ O(\epsilon^3) ,
\end{eqnarray}
where we have used
\begin{equation}
 \mbox{Tr}[-i(a-a^\dagger)\rho_{\mbox{\scriptsize c}}'] = 
 \mbox{Tr}[-i(\rho^{\mbox{\scriptsize a}}_{\mbox{\scriptsize 10}} -
 \rho^{\mbox{\scriptsize a}\dagger}_{\mbox{\scriptsize 10}})] 
+ O(\epsilon^3) .
\label{exptr}
\end{equation}
and defined
\begin{equation}
{\cal L}_{\mbox{\scriptsize m}}^{l} 
\rho^{\mbox{\scriptsize a}}_{\mbox{\scriptsize ij}} \; dt 
\equiv -\smallfrac{i}{\hbar}[H_{\mbox{\scriptsize m}} 
 - \hbar g (|\alpha|^2 + l) x, 
\rho^{\mbox{\scriptsize a}}_{\mbox{\scriptsize ij}} ] \; dt .
\end{equation}

In order to write a SME for the motion of the atom we need to
find a closed form equation for $\rho_{\mbox{\scriptsize a}} \approx
\rho^{\mbox{\scriptsize a}}_{\mbox{\scriptsize 00}} +
\rho^{\mbox{\scriptsize a}}_{\mbox{\scriptsize  
11}}$ but the differential equations for the diagonal elements of the
cavity mode density operator involve the off-diagonal elements. The
adiabatic elimination exploits the difference in time-scales between
the cavity and the motional dynamics by assuming that the heavily
damped  off-diagonal
elements have reached steady state values determined by the motional
state. This will allow the off-diagonal elements to be written in
terms of the diagonal elements which will result in the desired SME.
This
is a little more complicated than the usual adiabatic elimination,
because the off-diagonal elements are not merely strongly damped, but
also contain a stochastic driving term. 

To adiabatically eliminate
$\rho^{\mbox{\scriptsize a}}_{\mbox{\scriptsize 20}}$, we drop terms
proportional to $\rho^{\mbox{\scriptsize a}}_{\mbox{\scriptsize 20}}$
which are insignificant compared to the damping term, and obtain to
leading order 
\begin{eqnarray}
 d \rho^{\mbox{\scriptsize a}}_{\mbox{\scriptsize 20}} & = &  - \gamma 
 \rho^{\mbox{\scriptsize a}}_{\mbox{\scriptsize 20}} dt + 
 i\sqrt{2}g\alpha x \rho^{\mbox{\scriptsize a}}_{\mbox{\scriptsize 10}} 
 dt \nonumber \\
 & & + i \sqrt{\eta\gamma} \mbox{Tr}[ 
 \rho^{\mbox{\scriptsize a}}_{\mbox{\scriptsize 10}} 
 - \rho^{\mbox{\scriptsize a}\dagger}_{\mbox{\scriptsize 10}} ]
 \rho^{\mbox{\scriptsize a}}_{\mbox{\scriptsize 20}} dW .
\label{rho02}
\end{eqnarray}
For the purposes of showing that the contribution from the stochastic
driving is insignificant (ie. that it is not of leading order in
$\epsilon$), we assume now that $\rho^{\mbox{\scriptsize
    a}}_{\mbox{\scriptsize 10}}$ is constant. Since
$\rho^{\mbox{\scriptsize a}}_{\mbox{\scriptsize 10}}$ is actually
stochastically driven as well, this is not exactly correct; in the
steady state both off-diagonal elements will be fluctuating. However,
the resulting analysis demonstrates that these fluctuations are
higher order in $\epsilon$, so that setting $\rho^{\mbox{\scriptsize
    a}}_{\mbox{\scriptsize 10}}$ to its mean value is
self-consistent. With this
assumption Eq.(\ref{rho02}) is just linear multiplicative white
noise~\cite{Stochmeth}, with the stead-state solution
\begin{eqnarray}
 \langle \rho^{\mbox{\scriptsize a}}_{\mbox{\scriptsize 20}} 
 \rangle & = &  i\sqrt{2}\left(\frac{g\alpha}{\gamma}\right) x 
\rho^{\mbox{\scriptsize a}}_{\mbox{\scriptsize 10}} , \\
 \sigma^{\mbox{\scriptsize a}}_{\mbox{\scriptsize 20}} & = & 
  \sqrt{\smallfrac{\eta}{2}} |i\mbox{Tr}[ 
 \rho^{\mbox{\scriptsize a}\dagger}_{\mbox{\scriptsize 10}} - 
 \rho^{\mbox{\scriptsize a}}_{\mbox{\scriptsize 10}} ] \langle 
 \rho^{\mbox{\scriptsize a}}_{\mbox{\scriptsize 20}} \rangle | ,
\end{eqnarray}
where $\sigma^{\mbox{\scriptsize a}}_{\mbox{\scriptsize 20}}$ is the 
standard deviation of $\rho^{\mbox{\scriptsize a}}_{\mbox{\scriptsize 20}}$ 
in the steady-state. Since 
$\rho^{\mbox{\scriptsize a}\dagger}_{\mbox{\scriptsize 10}}$ is first order 
in $\epsilon$, the 
fluctuations about the steady-state are of third order in $\epsilon$, while 
the average value is of second order. We can therefore ignore the 
fluctuations to leading order in $\epsilon$, and write the result of 
the adiabatic elimination as 
\begin{equation}
  \rho^{\mbox{\scriptsize a}}_{\mbox{\scriptsize 20}} = 
i\sqrt{2}\left(\frac{g\alpha}{\gamma}\right) x 
\rho^{\mbox{\scriptsize a}}_{\mbox{\scriptsize 10}} + O(\epsilon^3) .
\label{re02}
\end{equation}
Proceeding in the same way to adiabatically eliminate 
$\rho^{\mbox{\scriptsize a}}_{\mbox{\scriptsize 10}}$, we find 
again that the fluctuations may be ignored to leading order, and obtain,
\begin{equation}
  \rho^{\mbox{\scriptsize a}}_{\mbox{\scriptsize 10}} = 
 2i\left(\frac{g\alpha}{\gamma}\right) 
 [ x \rho^{\mbox{\scriptsize a}}_{\mbox{\scriptsize 00}} 
- \rho^{\mbox{\scriptsize a}}_{\mbox{\scriptsize 11}} x ] + O(\epsilon^2),
\label{re01}
\end{equation}
Note that we have retained the term proportional to
$\rho^{\mbox{\scriptsize a}}_{\mbox{\scriptsize 11}}$, which is only
third order in $\epsilon$. However, retaining this term is useful as
this allows us to most easily recover the SME for the mirror. This
result (Eq.\ (\ref{re01})) confirms that $\rho^{\mbox{\scriptsize
    a}}_{\mbox{\scriptsize 10}}$ is indeed first order in
$\epsilon$. The fact that $\rho^{\mbox{\scriptsize
    a}}_{\mbox{\scriptsize 20}}$ is second order then follows
immediately from Eq.\ (\ref{re02}), and that $\rho^{\mbox{\scriptsize
    a}}_{\mbox{\scriptsize 11}}$ is second order follows from the fact
that $\rho^{\mbox{\scriptsize a}}_{\mbox{\scriptsize 10}}$ is first
order. Thus the assertion made above regarding the scaling of the
elements of $\rho_{\mbox{\scriptsize c}}$ is seen to be consistent
with the regime (\ref{assump1}). 

To obtain the stochastic master
equation we now substitute Eqs.\ (\ref{re02}) and (\ref{re01}) into
Eqs.\ (\ref{eq00}) and (\ref{eq11}) and combine them to obtain an
equation for $\rho_{\mbox{\scriptsize a}} \approx
\rho^{\mbox{\scriptsize a}}_{\mbox{\scriptsize 00}} +
\rho^{\mbox{\scriptsize a}}_{\mbox{\scriptsize   
11}}$. To leading order in $\epsilon$ the resulting stochastic master
equation for the atom is 
\begin{eqnarray}
  d\rho_{\mbox{\scriptsize a}} & = & -\frac{i}{\hbar}[H_m - \hbar
 g|\alpha|^2 x, \rho_{\mbox{\scriptsize a}} ] dt \nonumber \\ 
 & & + 2k{\cal D}[x] \rho_{\mbox{\scriptsize
a}} dt + \sqrt{2\eta k}{\cal H}[x] \rho_{\mbox{\scriptsize a}}dW   ,
\label{smepm}
\end{eqnarray}
This is the expected form for a process describing the
continuous measurement of position~\cite{cm1,cm2,cm3,cm4}. 
Note, however, that an extra term
proportional to $x$ appears in the effective Hamiltonian. This is just
the radiation pressure force on the mirror or the dipole force on the
atom. As this is
simply a classical linear force on the mirror, it may be cancelled by
applying an equal and opposite linear potential along with the trapping
potential, and we will assume that this is the case in all further
analysis. For the case of a measurement on an atom,
$k=2k_0^2g_0^4|\alpha|^2/(\gamma\Delta^2)$, while for the moving
mirror $k=2g_m^2|\alpha|^2/\gamma$. This quantity may be referred to
as the measurement constant, as it describes the rate at which information is
obtained about the atomic position, and the corresponding rate at which
noise is fed into the atomic momentum as a result of the measurement. 
Note that for unit efficiency detection and a pure initial state, the 
stochastic master equation is equivalent to a stochastic Schr\"{o}dinger 
equation for the state vector~\cite{cm1,cm2,cm4,cm7,WMhom}. 

The measurement signal is the photocurrent from the homodyne 
detection, being given by
\begin{equation}
  d\tilde{Q} = \beta [-\eta\gamma\langle i(a-a^\dagger) \rangle dt + 
\sqrt{\eta\gamma}dW] ,
\label{homsig}
\end{equation} 
where $\beta$ is determined by the strength of the local oscillator and 
the reflectivity of the beam splitter in the homodyne detection 
setup~\cite{WMhom}. 
Using Eqs.(\ref{exptr}) and (\ref{re01}) in Eq.(\ref{homsig}), we may 
write this as  
\begin{eqnarray}
  d\tilde{Q} & = & \beta [2\eta \sqrt{2\gamma k}\langle x \rangle dt 
 + \sqrt{\eta\gamma}dW] \nonumber \\
     & = & \beta\sqrt{\gamma/(2k)} [4\eta k\langle x \rangle dt 
 + \sqrt{2 \eta k}dW] ,
\label{possig}
\end{eqnarray}
and defining a scaled measurement signal by 
$dQ = d\tilde{Q}(\sqrt{2k/(\beta^2\gamma)})$, we may write 
\begin{equation}
  dQ = 4\eta k\langle x \rangle dt + \sqrt{2 \eta k}dW .
\end{equation}
The scaled photocurrent, $I(t) = dQ/dt$, may then be written as 
\begin{equation}
I(t) = 4\eta k\langle x \rangle + \sqrt{2 \eta k} \varepsilon(t),
\end{equation}
where $\varepsilon(t)$ is the delta correlated noise source 
corresponding to $dW$.

\section{Estimation and Feedback}
\label{solM}
We come now to the central part of this paper, the question of how to
employ a continuous measurement in the control of a quantum system. As
we noted in the introduction an arbitrary integral over the whole
photocurrent could be used to modulate an arbitrary feedback
Hamiltonian. This large number of degrees of freedom makes it
difficult to motivate any particular feedback scheme in real
systems. Perhaps as a result both theoretical and experimental efforts in
quantum feedback have focussed on feeding back the
photocurrent at each moment in time. In this paper we refer to this as
direct feedback. However, classical control theory~\cite{cfb,residual}
faces exactly the 
same problems and in this paper we propose a family of feedback
algorithms which is adapted from strategies employed in analogous
classical systems. Classical strategies often break the search for useful
control down into an estimation step and a control step. The powerful
technique of dynamic programming is able to find optimal algorithms
for both tasks in sufficiently simple systems~\cite{cfb}.
In general we propose that the estimation step for a quantum
mechanical system will involve the
solution in real time of an appropriate SME which accounts for
realistic levels of measurement and process noise, using an initially
mixed state 
reflecting lack of knowledge of the system. The state estimate which
results from the SME can then be used to modify the system Hamiltonian
in order to achieve the desired control of the system. 

In the previous section we reviewed how the stochastic master equation
(Eq.(\ref{smepm})), for the continuous position measurement of a 
single quantum degree of freedom may be derived from a real
measurement process. Fortunately, if the Hamiltonian for the mechanical
dynamics is no more than quadratic in the position and momentum, and the 
initial state of the system is Gaussian, the SME may be solved
analytically, since it remains Gaussian at all times~\cite{Jac1}. 
Taking the initial state to be Gaussian is also sensible, 
because there is reason to believe that non-classical states evolve
rapidly to Gaussians due to environmental interactions, of which
the measurement process is one example~\cite{gauss,Zurek}. A quantum
mechanical Gaussian
state is uniquely determined by its mean values and covariance matrix
(see for example~\cite{gardiner}),
just as is the case for classical probability distributions and so we
only need to find equations for these variables in order to fully
describe the evolution of the conditioned state. In what
follows, we will refer to the elements of the covariance matrix, 
being the position and momentum variance and their joint covariance,
simply as the {\em covariances}. The
expectation values of operators are found from 
\begin{equation}
d\langle c \rangle=\mbox{Tr}(cd\rho)
\end{equation}
as for any master equation, see for example~\cite{WandM}. 
The It\^{o} rules for stochastic differential 
equations~\cite{Stochmeth} result in equations for the covariances.
Performing this calculation gives the
equations for the means as~\cite{Doh1,means}
\begin{eqnarray}
d\langle x\rangle & = &  -\frac{i}{\hbar}\langle [x,H_m]\rangle dt  + 
2\sqrt{2\eta k}V_x \; dW, \nonumber \\ d\langle p\rangle & = & 
-\frac{i}{\hbar}\langle [p,H_m]\rangle dt  +  2\sqrt{2\eta k}C \; dW, 
\label{eqmns}
\end{eqnarray}
and the equations for the covariances are
\begin{eqnarray}
\dot{V}_x & = &  -\frac{i}{\hbar}\langle [x^2,H_m]\rangle  +
\frac{2i}{\hbar}\langle x\rangle\langle[x,H_m]\rangle -  8\eta k V_x^2,
\nonumber \\ 
\dot{V}_p & = &  -\frac{i}{\hbar}\langle [p^2,H_m]\rangle  +
\frac{2i}{\hbar}\langle p\rangle\langle[p,H_m]\rangle + 2k\hbar^2 -  
8\eta k C^2, \nonumber \\ 
\dot{C} & = & -\frac{i}{2\hbar}\langle [xp+px,H_m]\rangle  - 8\eta k V_x C
\nonumber \\ & & + \frac{i}{\hbar}\langle
x\rangle\langle[p,H_m]\rangle + \frac{i}{\hbar}\langle
p\rangle\langle[x,H_m]\rangle.
\label{eqcvs}
\end{eqnarray}
In these equations, $V_x$ and $V_p$ are the variances in position and 
momentum respectively, and
\begin{equation}
  C = \frac{1}{2}\langle xp + px\rangle - \langle x\rangle \langle p\rangle
\end{equation}
is the symmetrized covariance. The Gaussian assumption is
required to obtain Eqs.\ (\ref{eqcvs}) but not Eqs.\ (\ref{eqmns}). These
two systems of equations are precisely equivalent to the SME
(\ref{smepm}) under the assumption that the initial state is
Gaussian. 

First of all it should be noted that, while it is not explicit,
the equations for the covariances are closed, in that they do not depend
on the means, and, in addition, they do not depend upon the measurement
signal. As a consequence the covariances at any point in time depend only
upon the duration of the measurement, and not the specific measurement
record. These equations are instances of coupled Riccati equations, and
may be solved analytically~\cite{cm4,Doh1}. The full solutions are fairly
cumbersome, and we do not need to give them here. Once these equations have
been solved, the solutions may be substituted into the equations for the
means, and these are readily solved since they are merely linear
equations with (stochastic) driving. Writing them in the form
\begin{equation}
d \langle {\bf x} \rangle = A \langle {\bf x} \rangle dt +
2\sqrt{2\eta k}  \; d{\bf Y}(t)
\label{xp}
\end{equation}
where $\langle {\bf x} \rangle = (\langle x \rangle, \langle p \rangle)^T$
and $d{\bf Y}(t) = (V_x, C)^TdW$, the solution is naturally just
\begin{eqnarray}
\langle {\bf x} \rangle (t) = e^{At} \langle {\bf x} \rangle (0) + 
2\sqrt{2\eta k} e^{At}\int_0^t e^{-At'}  d{\bf Y}(t') .
\end{eqnarray}

During the measurement process two things happen. The first is that the 
mean position and momentum obey not only the evolution dictated by the
Hamiltonian, but also suffer continual random kicks due to the
measurement process. This is because, at each time step, the effect of the
measurement process is to perform a `weak' measurement of position, and
since the result of the measurement is necessarily random, the position
of the state in phase space changes in a random fashion~\cite{cm3}. 
The second effect of the measurement process, and the part that is
governed by the deterministic equations for the covariances, is to
narrow the width of the state in phase space. For ideal (unit efficiency)
detection, an initial mixture is reduced, over time, to a completely pure
state. 
At such a time there is, in that sense, no uncertainty, as the quantum
state is completely determined, and remains so. For inefficient
detection, the degree to which the state is mixed is also reduced during
the measurement, but, in general, to a non-zero level determined by the
detection efficiency~\cite{Doh1}.

Now let us consider how we might control the system by the process of
feedback. In classical control theory one attempts to obtain, at each
point in time, the best possible estimate of the state of the system, and
then uses the resulting estimates for the system variables in a feedback
loop to control the dynamics. Now, in the quantum example we are
considering here, since the distributions for all the variables are
always Gaussian, the mean position and momentum are also our {\em best
estimates} of these variables. In fact, it would be quite reasonable to
{\em define} a continuous measurement of a system variable as a process by
which we obtain an estimate of that variable at each point in time.
Hence, by this definition, what we need to do to achieve a continuous
measurement of a system variable, is to write down that integral of the
measurement signal which gives us continuously our best estimate of that
variable. The equations for the means are written above in terms of the
Wiener  process, rather than the actual measurement signal $dQ$.
Rewriting them in terms of the measurement signal we have
\begin{eqnarray}
d\langle x\rangle & = &  -\frac{i}{\hbar}\langle [x,H_m]\rangle dt  -
8\eta k \langle x\rangle V_x dt  + 2V_x dQ, \nonumber \\ d\langle
p\rangle & = &  -\frac{i}{\hbar}\langle [p,H_m]\rangle dt  - 8\eta k
\langle x\rangle C dt + 2C dQ, 
\label{eqsmns2}
\end{eqnarray}
so that the continuous position measurement does indeed provide us with a
continuous measurement of both position and momentum, in the sense
introduced above. The strategy we have outlined here of only employing the
mean values and not the variances of the conditioned state in the
control law turns out to be optimal in some classical systems
which are termed \emph{separable}.

The above equations require that we put in \emph{a priori} estimates of 
the state as the initial conditions.
While we expect the initial state to be highly mixed (so that in that
sense we initially have very little knowledge regarding the state), it
is assumed that one can obtain sensible estimates for the initial
means and covariances. This assumption is certainly reasonable, for
one can almost always obtain good estimates of the initial density
matrix from a knowledge of the way in which the system is
prepared. This is just as true in classical estimation theory. In that
case the initial values of the variables will be, in general, poorly
known, but good estimates for the initial probability distributions
for the variables (being analogous to the density matrix in the
quantum case) may be obtained from a knowledge of the initial
preparation. 

A further point to note is that after a sufficient time
the best estimates of the variables actually {\em do not depend upon
  the initial estimates}, but only upon the measurement record. Hence,
even an observer with very imprecise knowledge of the initial state
will have obtained accurate information after a time, and the
resulting feedback, while perhaps initially of no great advantage, will
eventually produce the desired effect. This property of the SME is
shown in 
reference~\cite{Doh1}, where the question of estimation is considered
in detail. As a final point regarding the question of estimation and
initial states, it worth noting that one can always wait a sufficient
time for the covariances to attain their steady-state values before
initiating feedback, thus obviating to a certain extent the need to
use initial estimates for the covariances. In classical control theory
this is concern about the errors in state estimates is termed {\em
  caution}, and in linear systems with 
quadratic costs and Gaussian noises caution turns out not to be optimal.

Curiously enough, Eqs.\ (\ref{eqsmns2}) do not admit of an analytic
solution  in 
terms of $dQ$ unless the covariances have their steady state
values. This is because the linear equation now has an explicit time 
dependence due to the fact that the time-dependent covariances multiply
the means. However, it is a simple matter to integrate these equations
numerically, and a computer could perform the necessary calculations to
obtain the best estimates of the variables in real time, and hence track
the evolution of the system. This process of estimation is not only
interesting because we can monitor the system evolution, but because the
estimates may be used in a feedback loop to control the dynamics.

Now that we know how to obtain the best estimates of the system
variables, the process of feedback involves continually adjusting the
Hamiltonian so that one or more of its terms are proportional to some
function of these estimates. In treating this process of feedback we must
be careful to ensure that the act of feedback (the act of adjusting the
Hamiltonian) happens after the measurement at each time step. This is
obviously essential, due to the fact that the measurement must be
obtained before any adjustment based on that measurement can take place.
However, in the limit of instantaneous feedback this is very simple.
First we consider the measurement step, in which the system evolves for a
time $dt$, and the measurement signal is incremented by the amount $dQ$.
At this point the feedback is allowed to act, and in the limit in which
it is instantaneous (that is, much faster than any of the time scales
which characterize the system dynamics), the Hamiltonian is updated
before the next time step. At the next time step the equations of motion
for the estimates (and, in fact, all system variables) have the new
Hamiltonian with the new values for the estimates, so that, effectively,
the Hamiltonian has the desired value at all times.  In the limit of
instantaneous feedback, the SME describing the
evolution of the system is therefore simply just as it was before, but
with the Hamiltonian, $H_m$, replaced with a new Hamiltonian, having
specific dependencies on the estimates of $x$ and
$p$, which are simply
$\mbox{Tr}[x\rho_{\mbox{\scriptsize a}}] $ and
$\mbox{Tr}[p\rho_{\mbox{\scriptsize a}}] $. The SME
describing the conditional evolution of the system, for general
instantaneous feedback via estimation from a continuous measurement of
position, is, therefore given by Eq.(\ref{smepm}), where $H_m$ is now a
function of the average position and momentum:
\begin{equation}
 H_m = f(x,p,\mbox{Tr}[x\rho_{\mbox{\scriptsize a}}] ,
\mbox{Tr}[p\rho_{\mbox{\scriptsize a}}] ) .
\label{gfh}\end{equation}
While the SME for the conditional evolution is 
therefore rather simple, particularly in that it is Markovian, an
equation that would describe the overall average (non-selective) evolution
would not be. This is because the average evolution at any given time is
not simply a function of the average density operator at that time, but
depends on the previous history. We have however provided a recipe to
calculate the 
unconditioned state at all times since this only requires averaging
over all the trajectories generated by our conditional feedback SME. 

We will now show that there is a precise analogy to be made between 
linear quantum mechanical systems (and in particular those subjected to 
continuous position measurement) and classical systems which are driven 
by a certain specified noise process. Once we define a quantum mechanical 
cost function, this precise analogy will allow us to identify the optimal 
feedback strategy by using classical LQG theory. 

In the estimation step of classical control, a system called a Kalman 
filter~\cite{cfb} is often used to obtain an estimate of the state of the 
system from the measurement record. Where the system is linear, and the noise
on the system (often referred to as {\em plant} or {\em process} noise) and 
the noise on the measurement are both Gaussian, this is the optimal state 
observer.
There is, in fact, a noise-driven classical system with noisy measurement
for which the Kalman filter turns out to be precisely Eqs.\ (\ref{eqmns}) 
and Eqs.\ (\ref{eqcvs}). Consider a classical harmonic oscillator with 
dynamical variables $x_{\mbox{\scriptsize c}}$ and 
$p_{\mbox{\scriptsize c}}$, obeying the equations~\cite{Doh1}
\begin{eqnarray}
  \dot{x}_{\mbox{\scriptsize c}} & = & p_{\mbox{\scriptsize c}}/m \\
  \dot{p}_{\mbox{\scriptsize c}} & = & -m\omega^2 x_{\mbox{\scriptsize c}}
+ \sqrt{2\eta k} \hbar \zeta_1(t) ,
\label{eqclas}
\end{eqnarray}
where the noise driving the classical system is delta correlated so that
$\langle \zeta_1(t)\zeta_1(t') \rangle = \delta(t-t')$, and the classical
measurement result, being also noisy, is given by 
\begin{equation}
  \dot{Q}_{\mbox{\scriptsize c}} = 4\eta k x_{\mbox{\scriptsize c}} +
\sqrt{2\eta k}\zeta_2(t) ,
\end{equation}
where $\langle \zeta_2(t)\zeta_2(t') \rangle = \delta(t-t')$ and
$\zeta_1$ and $\zeta_2$ are uncorrelated. The equations for the
best estimates and their covariances provided by the Kalman filter are
then exactly the same as for the quantum system, with the identification
\begin{equation}
dW = 2\sqrt{2\eta k} (x_{\mbox{\scriptsize c}} - \langle
x_{\mbox{\scriptsize c}}
\rangle)dt + \zeta_2 dt,
\end{equation}
which is referred to as the {\em innovation} or the {\em residual}. 
Hence, when the quantum states are Gaussian (and in that sense 
{\em classical}), the quantum 
measurement process may be viewed as a classical estimation process in
which noise, $\zeta_1(t)$, is continually fed into the system to maintain
the uncertainty relations. It is important to note that the strength of
the noise in this analogous classical process is determined by the
accuracy of the measurement. Unlike classical systems where
higher accuracy always results in better state estimation, high accuracy
in the position measurement will result in large
momentum variance, and hence large average energy, of the conditioned 
states. 

The existence of an analogous classical system is very nice because it 
allows us to use 
results from classical control theory when considering the quantum
system. In particular, when the cost function is quadratic in the system 
variables, a {\em separation theorem} applies to the classical system. 
This states that, given that it is the values of optimal estimates that 
being fed back, when calculating the feedback required for optimal 
control we may assume that the dynamical variables are known 
exactly~\cite{cfb,residual}: there is no advantage in considering the 
accuracy of our estimate. In this case the restriction of feeding back 
only best estimates is justified as leading to the optimal strategy. 
Moreover, another result which applies to the classical system states 
that the optimal control law will be the one which would be calculated
by assuming there is no noise either on the plant or on the measurement. 
This stronger property is termed certainty equivalence. All that remains 
to be done to allow us to match the quantum with the classical theory is 
to consider a class of quantum mechanical feedback Hamiltonians and a
quantum mechanical cost function, so as to complete the precise analogy 
between quantum and classical systems which exists for the Kalman filter 
and the SME.

We now examine briefly the relevant results from classical control theory. 
We note that classical optimal control theory has been applied in the 
past to closed (unmonitored) quantum systems by Rabitz and 
co-workers~\cite{Rabitz}. The classical system for which the Kalman filter 
is equivalent to the SME given by Eqs.(\ref{eqclas}) may be written 
\begin{equation}
 d {\bf x}_{\mbox{\scriptsize c}} = A {\bf x}_{\mbox{\scriptsize c}} +
\sqrt{2\eta k}(0,1)^T\zeta_1(t) dt + B {\bf u} ,
\end{equation}
where ${\bf x}_{\mbox{\scriptsize c}} = (x_{\mbox{\scriptsize c}},
p_{\mbox{\scriptsize c}})^T$ are the classical variables. Here we
have added feedback variables ${\bf u}$, which will be chosen to be some
function of the dynamical variables ${\bf x}_{\mbox{\scriptsize c}}$ so
as to implement control. Note that `optimal' control is defined as a 
feedback algorithm which minimizes a cost function, which is usually a 
function of the system state. The role of the cost function is to define 
how far the system state is from the desired state during the process of 
feedback-control. Classical LQG control theory tells us that for linear 
systems, driven by Gaussian noise, and in which the cost function is 
quadratic in the system variables (LQG stands for Linear, Quadratic, 
Gaussian), the optimal feedback is obtained by choosing 
${\bf u} = - K \langle {\bf x}\rangle$. The form of the cost function 
is chosen to be
\begin{equation}
 I = \int_o^t \left( {\bf x}_{\mbox{\scriptsize c}}^T P
 {\bf x}_{\mbox{\scriptsize c}} + {\bf u}^TQ{\bf u} \right) \; dt',
\end{equation}
and the optimal solution is the one that minimizes the expectation
value $J=\mbox{E}[I]$ of this cost function over the time that the
feedback acts. It turns out that $K=Q^{-1}B^TU$ and although $U$ is
time-dependent, the steady state value is often all that is used in 
practice, and obeys the equation~\cite{cfb} 
\begin{equation}
 0 = P + A^TU + UA - UBQ^{-1}B^TU .
\label{eqv}
\end{equation}
We will use these results in the next section when we consider cooling
a single quantum particle. Note that the choice of cost function is 
crucial in determining the optimal strategy. For example, placing 
boundaries on the available strength of feedback, rather than using a 
quadratic cost function typically implies that some form of bang-bang 
control (a control algorithm in which ${\bf u}$ takes one of two 
values) is optimal. In this case the optimal strategy is therefore not 
linear in the estimated state. 

So long as the feedback Hamiltonian given by Eq.\ (\ref{gfh}) is
linear in the position and momentum operators, so that it has the form
\begin{equation}
H_m = f(\mbox{Tr}[x\rho_{\mbox{\scriptsize a}}] ,
\mbox{Tr}[p\rho_{\mbox{\scriptsize a}}] ) x +
g(\mbox{Tr}[x\rho_{\mbox{\scriptsize a}}] ,
\mbox{Tr}[p\rho_{\mbox{\scriptsize a}}] ) p
\end{equation}
where $f$ and $g$ are arbitrary functions, then the dynamic equations for
the covariances {\em remain} decoupled from the equations for the means,
and remain deterministic. If this is not the case, then the
equations for the means become coupled to those for the covariances, and
the situation becomes more complex. 
Furthermore, if the feedback Hamiltonian has terms that are higher than
quadratic in $x$ and $p$ then the correspondence between quantum
feedback and LQG control will be lost since it will not preserve the
Gaussian property of the states and since the quantum Hamiltonian will
affect the state in a way that is distinct from the evolution of a
classical probability distribution. 

It remains to define a physically motivated quantum mechanical cost
function which maps onto the classical system we are considering. 
Clearly the part of the cost function which refers to the
performance of the system should be an expectation value of the 
unconditioned density operator. It should also be quadratic in the 
position and momentum, since this ensures that it is straightforward to
minimize a typical measure of control such as the average energy, just 
as it does for a classical system. On the other hand, when assessing the 
cost of control, it will be sufficient to consider classical quantities 
since the feedback Hamiltonian will typically be modulated by essentially 
classical quantities such as electric currents or lasers powers. With
these considerations, we can define a sensible quantum mechanical cost 
function as 
\begin{equation}
J_{q}=\int_{0}^{t}\left( \mbox{Tr}\left( {\bf x}^{T} P {\bf x} \rho 
\right) + \langle {\bf u}^{T} Q {\bf u} \rangle_{\mbox{\scriptsize
  c}}\right) \; dt 
\end{equation}
Where $\langle \rangle_{\mbox{\scriptsize c}}$ indicates an average
over the classical random variables ${\bf u}$. As noted above the 
density matrix average can be performed by first taking expectation 
values over the conditioned states given by the SME and then averaging 
over the trajectories. The cost function for a given trajectory is
\begin{equation}
I_{q}=\int_{0}^{t}\left( \langle {\bf x} \rangle ^{T} P \langle {\bf 
x} \rangle + \mbox{Tr} \left( PV\right) +{\bf u}^{T} Q {\bf u}\right)\; dt
\label{tcost}
\end{equation}
where the angle brackets indicate that we are talking about the mean
values and $V$ is the covariance matrix of the conditioned states. The 
second term in the integral is the expectation value of $P$ for the 
conditioned state if the mean ${\bf x}$ is zero. This term is independent 
of both the mean values and the feedback, so long as we use a linear 
feedback Hamiltonian as discussed above. It represents a minimum cost 
due to the finite width of the conditioned states. For a given trajectory, 
${\bf u}$ is not random, since it is a deterministic function of the 
current, and possibly of the past values of $\langle {\bf x}\rangle $. 
We can now re-express the quantum cost by averaging Eq.(\ref{tcost}) over 
the trajectories, which gives 
\begin{equation}
J_{q}=\int_{0}^{t}\left( \langle \langle {\bf x} \rangle^{T} P \langle 
{\bf x}\rangle \rangle_{\mbox{\scriptsize c}} + \mbox{Tr}\left( P V
\right) +\langle  
{\bf u}^{T} Q {\bf u} \rangle_{\mbox{\scriptsize c}}\right)\; dt.
\end{equation}
On the other hand, we have identified a classical problem for which
$\langle {\bf x} \rangle $ and $V$ obey the Kalman filter equations for 
the estimate of the noisy classical state 
${\bf x}_{\mbox{\scriptsize c}}$. Since the Kalman filter is merely 
sufficient statistics for a posterior probability distribution for 
${\bf x}_{\mbox{\scriptsize c}}$, we can write the average over
${\bf x}_{\mbox{\scriptsize c}}$ in $J$ in terms of this mean and 
covariance, giving
\begin{equation}
J=\int_{0}^{t}\left( \langle \langle {\bf x} \rangle^{T} P \langle
  {\bf x} \rangle \rangle_{c} + \mbox{Tr} \left( PV \right) 
  + \langle {\bf u}^{T} Q {\bf u} \rangle_{c}\right) dt .
\end{equation}
With this we see that the classical and quantum cost functions 
are identical. Thus the quantum 
mechanically optimal strategy (given that the feedback Hamiltonians are 
no more than quadratic) for the cost function we have introduced will
be the
classically optimal strategy for the fictitious classical system, whose
Kalman filter equations reproduce the SME, under the analogous
classical cost function. 
 Moreover, it is clear that for each 
SME describing a linear quantum system subjected to a linear 
measurement, there will be some classical LQG model which can be 
constructed to find the optimal feedback algorithm for a similarly
defined quadratic cost.

It is important to note that this is merely the optimal strategy for a 
given strength of measurement. For example, if the aim is to minimize 
the energy of an oscillator, overly strong position measurement will 
result in states of high average momentum and therefore energy. It is, 
however, straightforward to find the optimal measurement strength, 
if desired. One simply uses the procedure above to find the optimal 
strategy for a given measurement strength, $k$, and takes the extra 
step of optimizing the result over $k$. For the physical position 
measurements we discuss in section~\ref{cms}, changing the measurement
strength corresponds to changing the laser power driving the cavity. 

Up to this point we have not considered how particular feedback
Hamiltonians could be implemented, and so we complete this section with a
discussion of this important question. Clearly terms in the feedback
Hamiltonian proportional to functions of $x$ are implemented by applying
the required force to the system. By the use of estimation the
forces can be adjusted so that they are proportional to any particular
function of the average momentum and position as indicated above. This
allows terms to be added to the dynamical equation for the momentum, but
not to those for the position. We will show in the following section that
in order to achieve the best results for phase-space localization we must
add terms to the dynamical equation for the position, and therefore it is
important to be able to implement a term in the feedback Hamiltonian
proportional to momentum. This is not so straightforward, but we suggest
two possible ways in which it might be achieved. If the exact location of
the trap is not an important consideration, then shifts in the position
(being strictly equivalent to a linear momentum term in the Hamiltonian),
are achieved simply by shifting all the position dependent terms in the
Hamiltonian, in particular the trapping potential. This is a shift in the
origin of the coordinates, and, being a virtual shift in the position,
produces a term in the dynamical equation for the position proportional
to the rate at which the trap is being shifted. When the experimental
arrangement is such that the distance covered by the particle during the
cooling is negligibly small compared to the trapping apparatus this may
prove to be a very effective way of implementing a feedback Hamiltonian
linear in momentum. A second method would be to apply a large impulse to
the particle so that during one feedback time-step the particle is moved
the desired distance, and an equal and opposite impulse is then applied
to reset the momentum. Naturally the feasibility of this method will also
depend upon the practicalities of a given experimental arrangement.

\section{Cooling and Confinement via feedback}
\label{fb1}
\subsection{Using feedback by estimation}
Cooling and localization of individual quantum systems is an important
first step in the process of control. This is certainly true for trapped
atoms, ions, and cavity mirrors which we used as our examples in
section~\ref{cms}. By {\em cooling}, we mean localization in momentum
space, and by {\em confinement} we mean localization in position space.
When these two processes are combined, then we may speak of phase-space
localization. We now apply the formulation of quantum feedback introduced
in the previous section to the problem of phase space localization. As
indicated in that section, we can use classical control theory to find
the optimal feedback. First however, let us examine the steady state
solutions for the covariances in the absence of feedback. For a
harmonically trapped particle the equations for the covariances
become~\cite{cm4,Doh1} 
\begin{eqnarray}
\dot{V}_x & = & (2/m)C - 8k\eta V^2_x \\
\dot{V}_p & = & -2m\omega^2 C - 8k\eta C^2 + 2k\hbar^2 \\
\dot{C} & = & V_p/m - m\omega^2 V_x - 8k\eta CV_x
\label{cvhm}
\end{eqnarray}
The resulting steady-state covariances are
\begin{eqnarray}
V_x & = & \left(\frac{\hbar}{\sqrt{2\eta}m\omega}\right) 
          \frac{1}{\sqrt{\xi + 1}} \nonumber \\
V_p & = & \left(\frac{\hbar m\omega}{\sqrt{2\eta}}\right) 
          \frac{\xi}{\sqrt{\xi + 1}} \nonumber \\ 
C & = & \left(\frac{\hbar}{2\sqrt{\eta}}\right) 
          \frac{\sqrt{\xi - 1}}{\sqrt{\xi + 1}} .
\label{minvar}
\end{eqnarray}
where
\begin{equation}
\xi = \sqrt{1 + \frac{4}{\eta r^2}}, \;\; r = \frac{m\omega^2}{2\hbar\eta
k}
\end{equation}
Clearly the final state is in general a mixed Gaussian state, with the
exact orientation and squeezing determined by the measurement constant,
oscillation frequency, particle mass and detection efficiency. From these
covariances the purity of the final state is readily obtained by
using~\cite{Zurek}
\begin{equation}
  Tr[\rho_{\mbox{\scriptsize c}}^2] = 
(\hbar/2)(V_xV_p - C^2)^{-1/2}.
\label{eqfv}
\end{equation}
We find that the steady state purity of the monitored state is
\begin{equation}
  Tr[\rho_{\mbox{\scriptsize c}}^2] = \sqrt{\eta} .
\label{purity}
\end{equation}
For perfect detection efficiency ($\eta=1$), the state is therefore pure,
and perfectly determined at each point in time. For imperfect detection
the state is not completely pure, and is increasingly mixed as the
detection becomes less efficient. Inefficient detection also models
environmental noise, which in the case of a cavity mirror would be
coupling to a thermal bath, and in the case of a atom would be
spontaneous emission~\cite{Doh1}. The equations of motion for the
conditioned covariances are unchanged by linear feedback, whether direct
or using estimation, and Eq.\ (\ref{purity}) therefore gives the lower
limit on the purity of the final cooled state achievable for a given
detector efficiency.

Now we know the covariances and resulting purity of the conditioned
state, we want to know how well we can localize the mean position and
momentum of this state in phase space, by feeding back the estimated
values. The stochastic equations for the means are
\begin{eqnarray}
d\langle x\rangle & = &  (\langle p\rangle/m) dt  +  2\sqrt{2\eta k}V_x
dW, \nonumber \\  d\langle p\rangle & = & -m\omega^2 \langle x\rangle dt 
+  2\sqrt{2\eta k}C dW .
\end{eqnarray}
We wish to minimize the distribution of $\langle x \rangle$ and $\langle
p \rangle$ about the origin, and so it is sensible to take the cost
function to be minimized as in the previous section
\begin{equation}
J_{q}=\int_{0}^{t}\left( \mbox{Tr}\left( {\bf x}^{T} P {\bf x} \rho 
\right) + q^2 \langle {\bf u}^{T} Q {\bf u} \rangle_{\mbox{\scriptsize
  c}}\right) \; dt 
\end{equation}
where $q$ is a weighting constant which in this case has
units of time, and 
\begin{equation}
 P = Q = \left( \begin{array}{cc} m\omega^2 & 0 \\ 0 & 1/m \end{array}
\right) .
\end{equation}
With this choice of $P$ and $Q$ the cost function is a weighted sum of 
the energy of the oscillator and a fictitious energy one could 
associate with the feedback variable ${\bf u}$. In general one can 
choose any quadratic function of the feedback variables, and a 
particular choice would be made to suit a given situation. 
Note that the feedback variable ${\bf u}$ appears in the cost function 
to reflect the fact that we are not unrestricted in the magnitude of 
the feedback we bring to bear. If this consideration is relatively
unimportant, $q$ is chosen to be small, and so the cost function 
reduces essentially to the energy of the oscillator, which is 
certainly the quantity we wish to minimize in the process of 
phase-space localization.

With this form for the cost function, classical LQG control theory
tells us that linear feedback will provide optimal control. Choosing
$B=I$ in the feedback equation, so as to allow feedback in the
dynamical equations for both variables (the most general case), we need
merely solve Eq.(\ref{eqv}) for $V$ to find the optimal value of
the feedback matrix $K$. Performing this calculation we find that an
optimal solution is $K=(1/q)I$. That is, feedback to provide an equal
damping rate on both the position and momentum. Note that the smaller we
make the weighting constant $q$, the larger the damping rate, being
$\Gamma_x=\Gamma_p =\Gamma=1/q$. With this feedback, the dynamical
equations for the means become
\begin{eqnarray}
d\langle x\rangle & = & - \Gamma_x \langle x\rangle dt + (1/m)
\langle p\rangle dt  +  2\sqrt{2\eta k}V_x dW, \nonumber \\ 
d\langle p\rangle & = & - m\omega^2 \langle x\rangle dt - \Gamma_p \langle
p\rangle dt +  2\sqrt{2\eta k}C dW . \nonumber
\end{eqnarray}
It is now the mean and variance of the conditioned means which
are of interest, as they tell us how well localized the particle is, and
about what point in phase-space. We will denote the means of the
conditioned means as $\langle \langle x \rangle \rangle$ and $\langle
\langle p \rangle \rangle$, and the covariances of the means as
$V_x^{\mbox{\scriptsize e}}$, $V_p^{\mbox{\scriptsize e}}$ and
$C^{\mbox{\scriptsize e}}$, where the `e' refers to the fact that they
are {\em excess} to the quantum conditional covariances resulting from
the measurement process. Clearly the steady state values for the means
$\langle \langle x \rangle \rangle$ and $\langle \langle p \rangle
\rangle$ is the origin of phase space, while the equations for the
covariances are
\begin{eqnarray}
 \dot{\tilde{V}}_x^{\mbox{\scriptsize e}} & = & -2\Gamma_x
\tilde{V}_x^{\mbox{\scriptsize e}} + 2\omega \tilde{C}^{\mbox{\scriptsize
e}} + \frac{2\omega}{r} \tilde{V}_x^2 \\
 \dot{\tilde{V}}_p^{\mbox{\scriptsize e}} & = & -2\Gamma_p
\tilde{V}_p^{\mbox{\scriptsize e}} - 2\omega \tilde{C}^{\mbox{\scriptsize
e}} + \frac{2\omega}{r} \tilde{C}^2\\
 \dot{\tilde{C}}^{\mbox{\scriptsize e}} & = & -(\Gamma_x +
\Gamma_p)\dot{\tilde{C}}^{\mbox{\scriptsize e}} - \omega
(\tilde{V}_x^{\mbox{\scriptsize e}} - \tilde{V}_p^{\mbox{\scriptsize e}})
+ \frac{2\omega}{r} \tilde{C}\tilde{V}_x ,
\end{eqnarray}
and the tildes denote dimensionless scaled covariances given by
\begin{equation}
 \tilde{V}_x = \frac{2m\omega}{\hbar} V_x, \; \tilde{V}_p =
\frac{2}{\hbar m\omega} V_p, \; \tilde{C} = \frac{2}{\hbar} C .
\end{equation}
Putting $\Gamma_x=\Gamma_p=\Gamma$, and solving for the steady-state
covariances, we obtain 
\end{multicols}
\begin{eqnarray}
 \tilde{V}_x^{\mbox{\scriptsize e}} & = &
\frac{2{\cal Q}}{r(1+4{\cal Q}^2)} \left[ \left( 1 + 2{\cal Q}^2 \right)
\tilde{V}_x^2 + 2{\cal Q}^2 \tilde{C}^2 + 2{\cal Q} \tilde{C} \tilde{V}_x
\right] \\
 \tilde{V}_p^{\mbox{\scriptsize e}} & = & \frac{2{\cal Q}}{r(1+4{\cal
Q}^2)} \left[ 2{\cal Q}^2 \tilde{V}_x^2 + \left( 1 + 2{\cal Q}^2 \right)
\tilde{C}^2 - 2{\cal Q} \tilde{C} \tilde{V}_x \right] \\
 \tilde{C}^{\mbox{\scriptsize e}} & = & \frac{2{\cal Q}}{r(1+4{\cal
Q}^2)} \left[ -{\cal Q} \tilde{V}_x^2  + {\cal Q} \tilde{C}^2 + \tilde{C}
\tilde{V}_x
\right] ,
\end{eqnarray}
\begin{multicols}{2}
where ${\cal Q} \equiv \omega/(2\Gamma)$ .

The total average covariances resulting from the localization process
are simply the sum of the conditional covariances and these
excess covariances. The overall resulting purity may then be calculated
using Eq.(\ref{eqfv}), if so desired.

In the previous section we noted that terms in the feedback Hamiltonian
proportional to momentum are harder to generate than those proportional
to position, and since the optimal feedback we have used above requires
both kinds of terms, it is of interest to examine what may be achieved
with a position term alone. This imposes the condition that 
$K_{11}=K_{12}=0$, where $K_{ij}$ are the elements of the feedback matrix 
$K$. To derive the optimal solution under this condition we solve 
Eq.(\ref{eqv}) as before, but this time set
\begin{equation}
  B = \left( \begin{array}{cc} 0 & 0 \\ 0 & 1 \end{array} \right).
\end{equation}
This time taking the small $q$ limit, $q\omega=2{\cal Q}\ll 1$, an optimal 
feedback strategy is given by $K_{21}=m\omega/q$ and $K_{22}=1/q$.  In this 
case the steady state solution for the excess variances are
\begin{eqnarray}
 \tilde{V}_x^{\mbox{\scriptsize e}} & = & \frac{1}{r} \left[
 \tilde{V}_x^2 + 4{\cal Q}^2 \tilde{C}^2 + 4{\cal Q} \tilde{C} \tilde{V}_x 
 \right] \\
 \tilde{V}_p^{\mbox{\scriptsize e}} & = & \frac{1}{r} \left[
\tilde{V}_x^2 + 2 {\cal Q} \tilde{C}^2 \right] \\
 \tilde{C}^{\mbox{\scriptsize e}} & = & -\frac{1}{r} \tilde{V}_x^2 ,
\end{eqnarray}
We see, therefore, that using feedback by estimation, it is indeed
possible to obtain phase-space localization with only a position
dependent term in the feedback Hamiltonian, although this is clearly not
as good as using a combination of position and momentum damping.

To summarize our results so far, we see that when using feedback by
estimation, and when we average over the conditional evolution, there is
an additional uncertainty in the final localized state over that due to
measurement inefficiency, and that this excess uncertainty decreases with
the magnitude of the feedback. This additional uncertainty is due to the
noise which is continually fed into the system as the result of the
measurement. The effect of this noise is decreased as the damping constant, 
$\Gamma$, is increased. However, there is ultimately a limit upon the 
magnitude of the feedback, and hence upon $\Gamma$, and this is reflected 
in the choice of the weighting constant $q$ in the cost function. We will
see in the next section that direct feedback provides an alternative 
strategy for dealing with the measurement noise. 

\subsection{Adding direct feedback}

The beauty of direct quantum feedback, formulated by Wiseman and
Milburn~\cite{qfb}, is that it may be used to cancel the noise which
drives the mean values of the dynamical variables. This is possible
because the noise in the measurement signal is the {\em same} noise that
drives the system. Feeding back the measurement signal itself (by
choosing a feedback Hamiltonian directly proportional to this signal)
essentially allows the noise driving the system to drive it twice at each
step. If the feedback Hamiltonian is chosen in the right way, then the
effect of the noise at the first step may be canceled by that at the
second step. The result is that the steady state of the feedback
master equation can have the same variances as the conditioned states
since all of the fluctuations in the mean values are overcome.
We note that direct feedback would be analogous to the use of residual 
feedback in classical control theory (in which the innovation is fed back 
to drive the system) to cancel the noise driving the Kalman filter. 

As in the previous section, we choose the feedback Hamiltonian to be
linear in $x$ and $p$, as this is sufficient for our purposes. 
For direct feedback, the feedback Hamiltonian is proportional to the
measurement signal $I(t)$, so we may write
\begin{equation}
  H_D=I(t)(\alpha x + \beta p) . \label{fbHM}
\end{equation}
The stochastic master equation that results is~\cite{qfb}
\begin{eqnarray}
d\rho_{\mbox{\scriptsize a}} & = & -\frac{i}{\hbar}[ H_m, 
\rho_{\mbox{\scriptsize a}} ] dt + 2k{\cal D}[x] \rho_{\mbox{\scriptsize
a}} dt + \frac{1}{\eta}{\cal D}[F] 
\rho_{\mbox{\scriptsize a}} dt \nonumber \\
 &  & - i\sqrt{2k}[ F , x \rho_{\mbox{\scriptsize a}}  +
\rho_{\mbox{\scriptsize a}} x] dt \nonumber \\ 
 &  & + {\cal H}[\sqrt{2\eta k}x - \frac{i}{\sqrt{\eta}} F]
\rho_{\mbox{\scriptsize a}} dW  ,
\label{smedf}
\end{eqnarray}
where $F = (\sqrt{2 k}\eta)(\alpha x + \beta p)/\hbar$. This is
precisely the model that has been used previously to discuss the
manipulation of the motion of atoms and mirrors through feedback
\cite{Dun,mancini}. Applied 
to a harmonic oscillator and initially Gaussian states it is possible
to rewrite this master 
equation in terms of the mean values and covariances exactly as was done
above. 
The equations for the covariances are just as before
(Eqs.(\ref{cvhm})), but the equations for the means are now
\begin{eqnarray}
d\langle x\rangle & = &  (\langle p\rangle/m) dt  + 4\eta k \beta \langle
x \rangle dt +  \sqrt{2\eta k}(2V_x + \beta) dW, \nonumber \\  
d\langle p\rangle & = & -m\omega^2 \langle x\rangle dt - 4\eta k \alpha
\langle x \rangle dt +  \sqrt{2\eta k}(2C - \alpha) dW . \nonumber
\end{eqnarray}
We see that in order to cancel the noise driving the means we merely need
choose the feedback Hamiltonian such that $\alpha=2C$ and $\beta=-2V_x$.
However, it is also clear that direct feedback from a continuous position
measurement is limited in a way in which feedback by estimation is not.
Using direct feedback alone it is not possible to provide a damping
term for the mean momentum, or, in fact, any term in the equations for the
mean values which is proportional to the mean momentum. This is
because we are using continuous position measurement, so that the
measurement signal is proportional to the mean position, and not the mean
momentum. It is this limitation that feedback using estimation allows us
to overcome. Further, it is also clear that while feedback by estimation
allows us to achieve phase space localization even in cases where it was
not possible to provide the feedback Hamiltonian with a term involving
the momentum operator, direct feedback alone will not provide either 
cooling or confinement without the use of a momentum term. Consequently, 
a momentum term in the Hamiltonian is crucial for the cooling achieved in 
this system by the scheme of Mancini {\em et al.}~\cite{mancini}. 
Alternatively, in the absence of the momentum term, these equations are 
clearly well adapted to modifying the effective potential seen by the 
atom, and this is discussed at length by Dunningham 
{\em et al.}~\cite{Dun}.

It is important to note that the feedback Hamiltonian Eq.\ (\ref{fbHM}) 
could only be realized in the limit of an ideal (infinitely broad band) 
feedback signal, due to the fact that the measurement noise is white
noise  
(at least to an excellent approximation). The cost, $J_{q}$, of such a
signal  
is therefore also infinite. The use of linear direct feedback cannot 
improve the results obtained using feedback by estimation (since this is 
already optimal), and this is reflected in the fact that it eliminates the
noise only in the limit of infinite cost, a statement which is also
true of 
the optimal algorithm using feedback by estimation. However, direct 
feedback does provide an alternative strategy, and, depending on the
limitations imposed by a specific implementation, it might well prove
advantageous to use it in combination with feedback by estimation.   

To summarize, the lowest temperature available is given by the steady-state 
covariance matrix of the conditional states (Eqs.(\ref{minvar})), and in 
the limit in which the cost of control can be disregarded, either direct 
feedback or LQG control, or some combination (being analogous to classical 
LQG control plus residual feedback), give a means of achieving this as a 
limiting case. The result is that the final cooled, localized state has the 
covariances given by Eqs.(\ref{minvar}), and the resulting purity, given 
by Eq.(\ref{purity}), would be limited only by the detection efficiency 
and environmental noise.

\section{Conclusion}
\label{conc}
In this paper we have shown that it is possible to formulate, in a simple
manner, feedback in linear quantum systems such that the best estimates
of system variables are used to control the system. This
significantly extends the range of available possibilities for the
control of quantum systems using feedback. Due to the fact that in
linear systems the estimation process may be modeled by its classical 
analogue, Kalman filtration, classical LQG control theory may be applied 
to quantum feedback by estimation.

While we have focused on applying results from LQG theory to linear
systems, there are many other techniques from classical state observer
based control which could be applied to control quantum systems. For
example, the techniques of adaptive control, where system parameters are
estimated on-line in order to cope with non-linearities or
uncertainties about the system to be controlled, could be employed. 
Another problem to be faced in more complicated systems is the
computational overhead in propagating the state estimate. Linear
systems are tractable classically because the mean and covariance
matrix provide all the necessary information about the posterior 
probability distribution with the result that the whole distribution 
need not be propagated. Propagating the SME's of non-linear quantum 
systems in real time will require extensive computational resources. 
Approximations to the full SME, perhaps along the lines of classical 
extended Kalman filters, may well be useful or necessary in real 
near-future experiments. 

\section*{Acknowledgments}
ACD would like to thank Hideo Mabuchi, Sze Tan and Dan Walls for helpful
discussions, and the University of Auckland for financial support. KJ 
would like to thank Prof. Dan Walls for hospitality during a stay at
the University of Auckland where this work was carried out. 

\end{multicols}

\vspace{-0.2cm}
\begin{multicols}{2}

\end{multicols}

\end{document}